\documentclass[runningheads]{llncs}
\usepackage[T1]{fontenc}
\usepackage{color}
\usepackage{algorithm}
\usepackage{algorithmic}
\usepackage{graphicx}
\usepackage{amsmath}
\usepackage{amsfonts}
\usepackage{multirow}
\usepackage{balance} 
\usepackage{mathrsfs}
\usepackage{makecell}
\usepackage{booktabs,array}

\begin{document}
\title{Large Language Model Can Be a Foundation for Hidden Rationale-Based Retrieval}
\titlerunning{Large Language Model for Hidden Rationale-Based Retrieval}
%
\author{Luo Ji\inst{1}\orcidID{0000-0002-2484-5345} \and
Feixiang Guo\inst{1}\thanks{\;The first two authors contributed equally to this research.} \and
Teng Chen\inst{1}\orcidID{0009-0000-3604-5313} \and
Qingqing Gu\inst{1} \and
Xiaoyu Wang\inst{2} \and
Ningyuan Xi\inst{3}\orcidID{0009-0008-7529-4660} \and
Yihong Wang\inst{1} \and
Peng Yu\inst{1} \and
Yue Zhao\inst{1} \and
Hongyang Lei\inst{1} \and
Zhonglin Jiang\inst{1} \and
Yong Chen\inst{1}\thanks{\;Corresponding author.}
}
\authorrunning{L. Ji et al.}
%
\institute{Geely AI Lab,
\\ \email{\{Luo.Ji1,Feixiang.Guo1,Teng.Chen2,Qingqing.Gu3,
YiHong.Wang2,Peng.Yu15,Yue.Zhao17,Hongyang.Lei,
zhonglin.jiang,yong.chen\}@geely.com}
\and
Beijing Institute of Technology,
\email{3220230388@bit.edu.cn}\\
\and
Beihang University,
\email{21373102@buaa.edu.cn} 
}
\maketitle              
\begin{abstract}
Despite the recent advancement in Retrieval-Augmented Generation (RAG) systems,  most retrieval methodologies are often developed for factual retrieval, which assumes query and positive documents are semantically similar. In this paper, we instead propose and study a more challenging type of retrieval task, called hidden rationale retrieval, in which query and document are not similar but can be inferred by reasoning chains, logic relationships, or empirical experiences. To address such problems, an instruction-tuned Large language model (LLM) with a cross-encoder architecture could be a reasonable choice. To further strengthen pioneering LLM-based retrievers, we design a special instruction that transforms the retrieval task into a generative task by prompting LLM to answer a binary-choice question. The model can be fine-tuned with direct preference optimization (DPO). The framework is also optimized for computational efficiency with no performance degradation. We name this retrieval framework by \textbf{RaHoRe} and verify its zero-shot and fine-tuned performance superiority on Emotional Support Conversation (ESC), compared with previous retrieval works. Our study suggests the potential to employ LLM as a foundation for a wider scope of retrieval tasks.  Our codes, models, and datasets are available on \url{https://github.com/flyfree5/LaHoRe}.

\end{abstract}
%

%

\section{Introduction}

\begin{figure*}[!htbp]
  \includegraphics[width=1\linewidth]{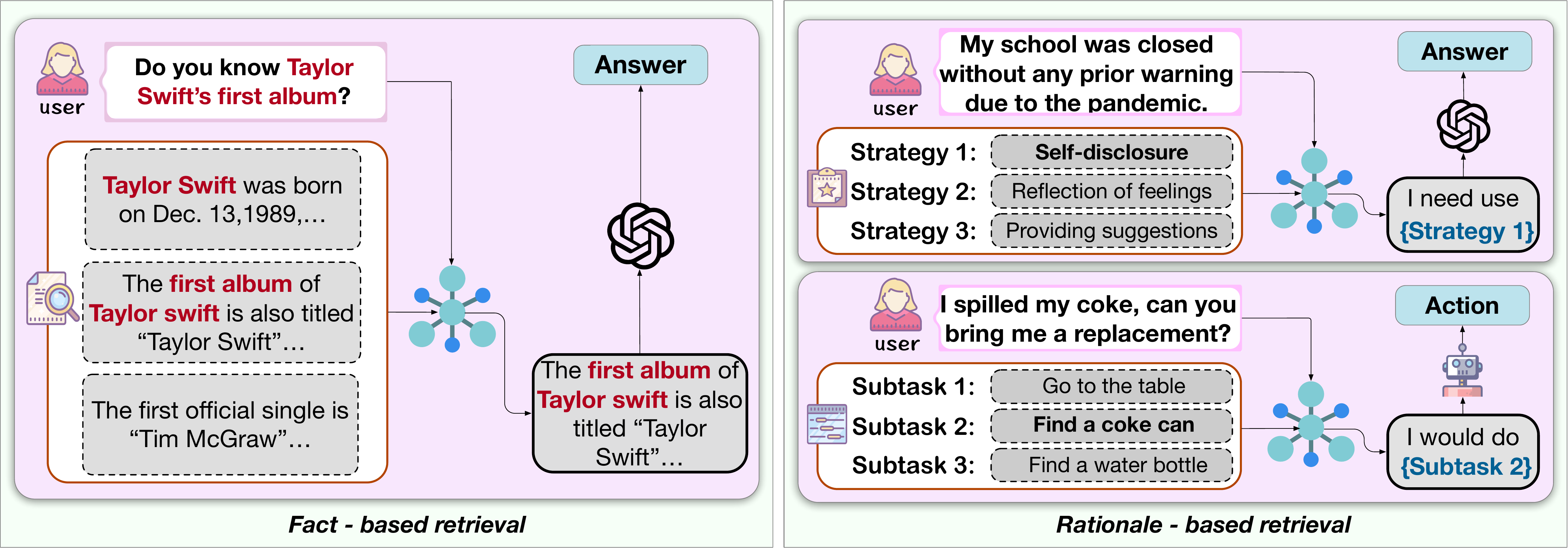} 
  \caption {Comparison of retrieval paradigms. Conventional fact-based retrieval tasks (Left) can be studied by query-document semantic similarity, while rationale-based retrieval tasks (Right) such as reply strategy or embodied subtask retrieval can not.}
  \label{fig:paradigm}
\end{figure*}

Retrieval-Augmented Generation (RAG) has demonstrated remarkable progress based on the development of the Large Language Model (LLM), which combines textual retrieval and data-augmented generation, to improve the response's professionalism, controllability, explainability, and reduce hallucination \cite{gao2024retrievalaugmentedgenerationlargelanguage,zhao2024retrievalaugmentedgenerationaigeneratedcontent,fan2024RAGmeetingLLMs,zhao2024retrievalaugmentedgenerationrag}. Given the user query, a retriever is obliged to select the most relevant document from the document corpus, to prepend into the generated prompt. Relevance between query and document can usually be drawn from semantic textual similarity (STS), in which the frequency-based retrieval methods \cite{Robertson2009BM25}, or bi-encoder architecture of dense retrievers \cite{asai2023tart,su2023instructorXL,wang2024E5} can be applied. Such retrieval methodologies have been proven to be quite effective in traditional factual or knowledge-based question-answering systems (Fig.~\ref{fig:paradigm} (Left)).




However, wide applications of RAG systems sometimes challenge LLM for higher levels of cognitive capabilities, in which query and document are semantically related but not paraphrases of each other \cite{wang2024E5}, and retrieval needs to be inferred from hidden rationale instead of explicitly factual-based \cite{zhao2024retrievalaugmentedgenerationrag,yang2024BPO}. For example, one can develop a strategy pool and build a user-centric dialogue system that retrieves an appropriate response strategy given the user query \cite{zhao-etal-2023-transesc,cheng-etal-2022-improving,qiu2024psychatclientcentricdialoguemental}; a robot can search for the most valuable sub-task from plausible candidates determined by LLM, grounded by a global task instruction \cite{saycan2022arxiv,huang2022inner,knowno2023}  (Fig.~\ref{fig:paradigm} (Right)). Although such problems can be viewed as or implemented with retrieval-based systems, traditional similarity-based retrievers might fail, since the relevance reasoning now needs to be built upon underlying logical congruence or thematic alignment. Instead, LLMs with inner-embedded and in-context commonsense knowledge might be crucial tools, especially with external data insufficiency. 

\begin{figure*}[!htbp]
  \includegraphics[width=1\linewidth]{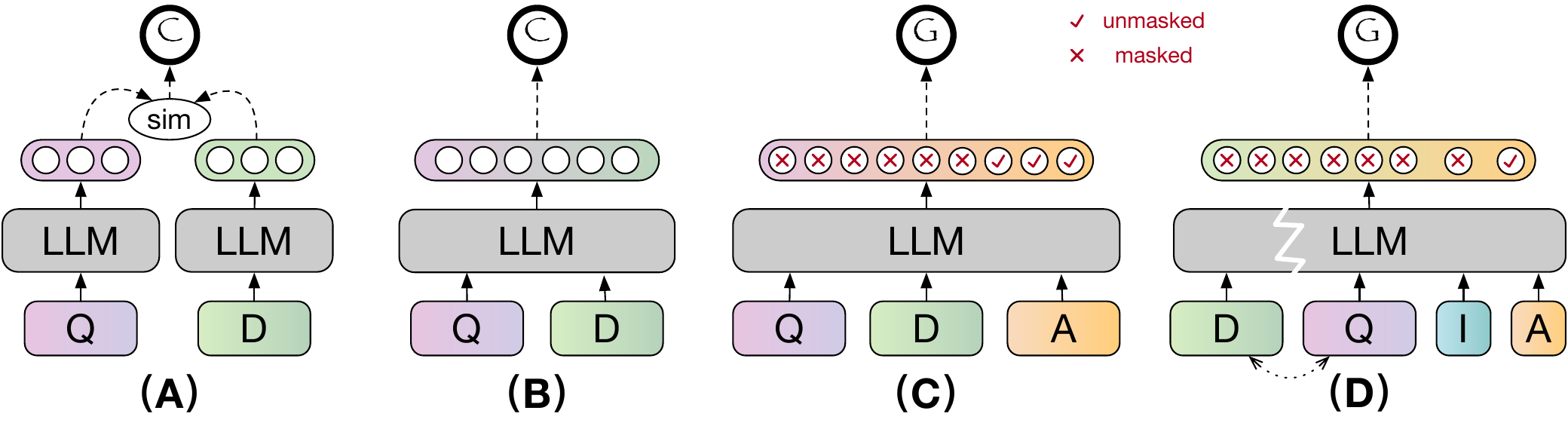}
  \caption {Architectural Comparison of {LaHoRe} with previous LLM-based retrieval methods. C, G, sim denote contrastive loss, generative loss, similarity function, respectively.\\ 
  (A): Bi-encoder; (B): Cross-encoder; (C): Generative RAG framework; (D): Framework of {LaHoRe}. We switch the order of Q and D such that D's encoding can be cached.  }

  \label{fig:framework}
\end{figure*}

Recent efforts of LLM-based retrieval either keep the bi-encoder architecture \cite{ma2024RepLLaMA,weller2024promptrieverinstructiontrainedretrieversprompted,wang2024E5} (Fig.~\ref{fig:framework} (A)), or processes the query-document concatenation by cross-encoding \cite{muennighoff2022sgpt,li2024llama2vec} (Fig.~\ref{fig:framework} (B)), or even combined with generative loss or task \cite{muennighoff2024GritLM,zhang2024onegen} (Fig.~\ref{fig:framework} (C)). However, most of these efforts are still embedding-based approaches, with the representing learning component achieved by contrastive loss of in-batch negatives or hard negatives \cite{li2024matchinggenerationsurveygenerative}. Inspired by recent successful application s of generation tasks on retrieval studies, it is argued that an LLM-based cross-encoder with a solely generative loss might be able to understand the non-explicit semantic relation between query and document \cite{nguyen2024sfrrag}, becoming a foundation for hidden rationale-based retrieval problems.





In this paper, we propose a novel approach called \textbf{L}LM-b\textbf{a}sed \textbf{H}idden Rati\textbf{o}nale-based \textbf{Re}triever (\textbf{LaHoRe}), to address the aforementioned challenges. We combine the ideas of cross-encoder and the generative architectures, to conduct a paradigm shift from discriminative to generative modeling (Fig.~\ref{fig:framework} (D)). An instruction is appended to prompt a binary-choice question on the retrieval relevance, while the relevance score can be extracted from the choices' relative log probability. To validate the effectiveness of {LaHoRe}, we construct several datasets from the field of Emotional Support Conversation (ESC) \cite{liu2021ESconv} and verify that {LaHoRe} outperforms previous baselines on the supporting strategy retrieval. We also utilize the prompt cache technology to improve the computational efficiency by changing the order of query and document. The zero-shot performance of {LaHoRe} can be further improved by SFT or DPO. The main contributions include the following:
\begin{itemize}
    \item We expand the scope of traditional retrieval with a more challenging category called hidden rationale-based retrieval, and experiment it in the scenario of emotional support conversations.
    \item We design a pure generative-based cross-encoder framework and convert the conventional contrastive loss into preference learning, with state-of-the-art performance obtained.
    \item We propose a special solution to obtain a computationally efficient retriever.
\end{itemize}

\section{Method}

In this section, we first introduce our prompt format, then propose the inference logic that provides zero-shot results. Finally the fine-tuning approach is discussed which further improves the performance. Fig.~\ref{fig:framework} (D) visualizes our framework.

\subsection{Prompt Format}

Given a query $Q$ and a document $D$, we design a special instruction, named $I$, appended to the end of query-document concatenation:
\begin{equation}
\label{eq:input}
    \text{input} = \text{document: } \{ D \} \backslash n \text{ query: } \{ Q \} \backslash n \text{ } \{ I \} 
\end{equation}
The purpose of instruction is to transform the retrieval task into a generation task, which contains two parts, $\{ I \} = \{ I_r \} 
 \backslash n  \{ I_{\text{bc}} \}$:

\noindent $\{ I_r \}$ = \textit{"Can Q be appropriately responded with D?"} \\
\noindent $\{ I_{\text{bc}} \}$ = \textit{"If you think the answer is true, choose <T>; otherwise choose <F>."} \\
where $\{ I_r \}$ connects $D$ and $Q$ from the sense of semantic relation, and $I_{\text{bc}}$ encourages LLM to thoroughly consider its preference on the relevance by augmenting a binary-choice problem \cite{Zhang2024R-Tuning}, and finally provide the answer $A$. An instruction fine-tuned LLM can constrain its $A$ by ("<T>", "$<F>$"), as expected.



\subsection{Inference}
\label{sec:inference}

The relevance score is deduced from the positive-negative relative confidence:
\begin{align}
    s(A) = \log \left[ \text{P}(A|\text{input}) \right], \quad s_{\text{rel}} = s(<T>) / (s(<T>) + s(<F>)) \label{eq:inference} 
\end{align}
where $\log \left[ \text{P}(\cdot) \right]$ denotes LLM's next-token log probability. Although simple, we find this pipeline performs well even on zero-shot tests. However, it is computationally expensive since it requires a full forward pass for each $(Q, D)$ pair.


\noindent \textbf{Conversion into a efficient retriever: } To alleviate computational complexity, we intentionally put $D$ before $Q$ in Equation \ref{eq:input}, namely $D \Rightarrow Q$, instead of the more semantically natural order ($Q \Rightarrow D$). Given the causal decoding mechanism of LLM, the prefix decoding or cache-sharing technology \cite{kwon2023vLLM} can be utilized to calculate and store the key-value cache of each $D$ off-the-shelf, while only decoding the part of $(Q, I)$ during real-time inference with the cache loaded:
\begin{align*}
  \text{offline: } \text{cache}_{KV}(D) = \text{LLM}(D), \text{online: } s(A) = \log \left[ \text{P}(A|\text{cache}_{KV}(D), Q, I) \right]
\end{align*}


\subsection{Annotation and Fine-Tuning}
\label{sec:training}

To leverage the scaling capability of LLM, each query is combined with all mismatched documents from the pool, to create the negative pairs. In the experiment section, we call this positive-negative ratio by `1:all'. The labels of $A$ are then annotated with "<T>" for positive pairs and "<F>" for negative pairs.

A post-hoc adaption on the instruct LLM can be conducted either by Supervised Fine-Tuning (SFT) \cite{Ouyang2022ChatGPT} or Direct Preference Optimization (DPO) \cite{Rafailov2023DPO}, where DPO can be viewed as the generative version of the contrastive loss.
 



\section{Experiment}


In this section, we first introduce the datasets we use, then list zero-shot and fine-tuning results. Ablation and sensitivity studies are also conducted.

\begin{table*}[t!]
    \caption{An example of \textit{ESconv} as a retrieval dataset. `All Strategies' corresponds to the document list, and `Strategy' is the ground truth of retrieved document.}
    \label{tab:ex_esconv}
    \centering
    \resizebox{\textwidth}{!}{
    \begin{tabular}{c|l}
        \toprule
        All Strategies & \makecell[l]{[Question,Restatement or Paraphrasing,Reflection of feelings,Information, \\
        Self-disclosure,Affirmation and Reassurance,Providing Suggestions,Others]}  \\
        \toprule
        Query & \makecell[l]{\textit{\{history\}} \\
        \textit{user:} Seriously! What I  am scare of now is how to secure another job.}  \\
\midrule
       Strategy & \makecell[l]{\textcolor{blue}{Reflection of feelings}}   \\
\midrule
      Response & \makecell[l]{\textit{assistant:} I can feel your pain just by chatting with you.}   \\
        \bottomrule
    \end{tabular}
    }

\end{table*}

\subsection{Datasets}



Experimental datasets are constructed from either proprietary or open sources:

\noindent \textbf{Emotional Support Conversation (ESC): } we utilize two public datasets, \textit{ESconv} \cite{liu2021ESconv} and \textit{PsyQA} \cite{sun-etal-2021-psyqa}, with `support strategy' annotated for each assistant utterance. \textit{ESconv} is multi-turn with 1202 dialogues, 12746 utterances, and 8 strategies. \textit{PsyQA} is single-turn with 4012 dialogues and 7 strategies.

\noindent \textbf{Daily Chat: } we collect a proprietary multi-turn daily chat dataset including 1855 dialogues and 7910 utterances. Each turn is annotated with a user-side intent (\textit{e.g.}, `user asks for help' or 'user disagrees') and an assistant-side reply strategy (\textit{e.g.}, `I need to show empathy' or 'I should ask user's habit') by human labelers. Two retrieval datasets called \textit{ChatIntent} and \textit{DailyStrategy} are then created with 98 intents and 82 strategies totally. 

For all four datasets, each user utterance is prepended by the historical context (if any) to form the query, with all possible strategies or intents as documents. One can refer to Table \ref{tab:ex_esconv} for an exampled sample from \textit{ESconv}. \textit{PsyQA} and \textit{DailyStrategy} have similar formats while \textit{ChatIntent} has the intents as documents instead of strategies.


\begin{table*}[h!]
\centering
\caption{Zero-shot performances on full datasets. `R' denotes Recall. Bold indicates the best result while underline indicates the second best. Numbers are in percentage.}
\label{tab:zero_shot_result}
\resizebox{\textwidth}{!}{%
\begin{tabular}{ll|c|ccc|ccc|ccc|ccc}
\toprule
 & \multicolumn{2}{c|}{Dataset($\rightarrow$)} & \multicolumn{3}{c|}{ChatIntent} & \multicolumn{3}{c|}{ChatStrategy} & \multicolumn{3}{c|}{ESconv} & \multicolumn{3}{c}{PsyQA}  \\
\cline{2-3}  \cline{4-6} \cline{7-9} \cline{10-12} \cline{13-15}
 & Method ($\downarrow$) & Size & R@5 & R@10 & p-MRR & R@5 & R@10 & p-MRR & R@1 & R@3 & p-MRR & R@1 & R@3 & p-MRR \\
\midrule
\parbox[t]{5mm}{\multirow{8}{*}{\rotatebox[origin=c]{90}{
\parbox[c]{2cm}{\centering \scriptsize Bi-Encoder}}}} 
& BM25\cite{Robertson2009BM25} & - & 5.4 & 11.2 & 5.6 &6.2 & 11.4 & 5.7 & 7.3 & 30.4 & 31.6 & \bf 33.3 & \underline{55.2} & \underline{48.2} \\
& TART-Contriever\cite{asai2023tart} & 110M & 4.2 & 9.2 & 5.1 & 4.8 & 8.9 & 5.1 & 1.4 & 38.5 & 35.1 & 11.3 & 44.2 & 36.5 \\
& E5-large-instruct\cite{wang2024E5} & 550M & 28.2 & 40.0 & 21.1 & 11.0 & 18.9 & 9.4 & \underline{16.4} & \underline{45.4} & \bf 39.0 & 16.3 & 48.1 & 39.9 \\
& InstructOR XL\cite{su2023instructorXL} & 1.5B & 2.9 & 6.8 & 5.2 & 5.4 & 10.4 & 6.9 & 13.7 & 44.3 & 36.4 & 24.6 & \bf 61.2 & 48.1 \\
& RepLLaMA\cite{ma2024RepLLaMA} & 7B & 14.9 & 23.5 & 11.8 & 6.2 & 11.6 & 6.2 & 11.5 & 40.2 & 33.9 & 26.8 & 47.1 & 45.6 \\  
& Promptriever\cite{weller2024promptrieverinstructiontrainedretrieversprompted} & 7B & 22.2 & 32.7 & 17.3 & 7.0 & 13.3 & 6.5 & 12.0 & 42.7 & 35.4 & 25.3 & 49.1 & 45.0 \\ 
& E5-Mistral\cite{wang2024E5} & 7B & 27.6 & 39.5 & 19.9 & 9.8 & 18.3 & 9.5 & 11.5 & 40.5 & 34.6 & 12.5 & 45.6 & 37.2 \\
 \midrule
\parbox[t]{5mm}{\multirow{3}{*}{\rotatebox[origin=c]{90}{
\parbox[c]{1.2cm}{\centering \scriptsize Cross-Encoder}}}}
& LlaMA2Vec\cite{li2024llama2vec} & 7B & 15.9 & 26.6 & 12.0 & 4.8 & 10.0 & 5.4 & \underline{16.4} & 44.4 & 38.0 & 28.0 & 49.4 & 46.7 \\ 
& GritLM\cite{muennighoff2024GritLM} & 7B & \underline{29.5} & \underline{40.2} & \bf 22.3 & \underline{14.4} & \underline{23.9} & \underline{11.0} & 15.9 & \bf 45.8 & \underline{38.5} & 5.3 & 37.9 & 30.8 \\ 
\cline{2-15}
 & \textbf{{LaHoRe}} (ours) & 7B & \bf 30.7 & \bf 48.2 & \underline{22.1} & \textbf{16.5} & \textbf{28.3} & \textbf{11.6} & \textbf{16.9} & 42.7 & 38.0 & \underline{32.6} & 52.5 & \textbf{49.5} \\
\bottomrule
\end{tabular}
}
\end{table*}

\subsection{Results}

The train-test split of datasets is 9:1. Considering the amount of documents, here we examine Recall@5, Recall@10, p-MRR for \textit{ChatIntent} and \textit{DailyStrategy}, and Recall@1, Recall@3, p-MRR for \textit{ESconv} and \textit{PsyQA}. LaHoRe is initialized from Qwen2-7B-Instruct \cite{qwen2techreport2023}. Training is conducted on LlamaFactory \cite{zheng2024llamafactory}, with the learning rate of $1.0e-6$, window length of 2048 and epoch of 3. Batch size is set as 512 for SFT and 128 for DPO. 


%


\begin{table*}[t!]
\centering
\caption{Fine-tuning performances on test sets. `R' denotes Recall. Bold indicates the best result while underline indicates the second best. Numbers are in percentage.}
\label{tab:finetune_result}
\resizebox{\textwidth}{!}{%
\begin{tabular}{ll|cccc|cccc|ccc|ccc}
\toprule
& \multirow{1}{*}{Dataset($\rightarrow$)} & \multicolumn{4}{c|}{ChatIntent} & \multicolumn{4}{c|}{ChatStrategy} & \multicolumn{3}{c|}{ESconv} & \multicolumn{3}{c}{PsyQA}  \\
\cline{2-2}  \cline{3-6} \cline{7-10} \cline{11-13} \cline{14-16}
& Method ($\downarrow$) & R@1 & R@5 & R@10 & p-MRR & R@1 & R@5 & R@10 & p-MRR & R@1 & R@3 & p-MRR & R@1 & R@3 & p-MRR \\
\midrule
\parbox[t]{5mm}{\multirow{3}{*}{\rotatebox[origin=c]{90}{
\parbox[c]{1.2cm}{\centering \scriptsize Bi-Encoder}}}} 
& RepLLaMA & 8.2 & 22.7 & 37.3 & 17.2 & 4.1 & 20.0 & 29.3 & 13.4 & 17.4 & 49.1 & 39.7 & 29.6 & 56.0 & 49.6 \\ 
& Promptriever & 8.4 & 23.0 & 35.2 & 18.0 & 4.1 & 23.2 & 31.8 & 14.3 & 17.6 & 48.9 & 39.5 & 29.6 & 61.7 & 50.9 \\   
& E5-Mistral & 31.1 & 59.5 & 71.8 & 44.9 & 28.4 & 61.1 & 74.8 & 43.8 & 15.5 & 50.2 & 39.3 & 11.4 & 48.0 & 36.8 \\ 
\midrule
\parbox[t]{5mm}{\multirow{4}{*}{\rotatebox[origin=c]{90}{
\parbox[c]{1.6cm}{\centering \scriptsize Cross-Encoder}}}} 
& LlaMA2Vec & 35.7 & 71.1 & 80.7 & 50.6 & 37.5 & 76.4 & 85.0 & 53.8 & 23.9 & 55.6 & 45.5 & 34.3 & 75.4 & 57.5 \\ 
& GritLM & 28.2 & 60.7 & 71.8 & 42.8 & 29.8 & 66.4 & 82.5 & 46.8 & 19.2 & 50.2 & 41.7 & 7.7 & 32.1 & 29.8 \\  
\cline{2-16}
& \textbf{{LaHoRe}}-SFT & \underline{64.1} & \underline{87.3} & \underline{93.6} & \underline{73.4} & \underline{61.4} & \underline{88.9} & \underline{95.0} & \underline{75.1} & \underline{34.4} & \underline{67.1} & \underline{54.8} & \underline{35.6} & \textbf{77.6} & \underline{58.2} \\
& \textbf{{LaHoRe}}-DPO & \bf 77.3 & \bf 93.6 & \bf 95.9 & \bf 83.8 & \bf 73.9 & \bf 94.1 & \bf 97.5 & \bf 82.4 & \bf 38.0 & \bf 69.5 & \bf 57.6 & \textbf{35.8} & \underline{76.6} & \textbf{58.4} \\
\bottomrule
\end{tabular}
}
\end{table*}

\noindent \textbf{Zero-shot Evaluation: } 
Table \ref{tab:zero_shot_result} compares zero-shot results between Bi-Encoder and Cross-Encoder architectures. LLM-based baselines do not perform well and are even worse than smaller retrievers in some metrics. Our LaHoRe performs the (or the second) best across all metrics, which indicates LaHoRe can utilize knowledge and reasoning of LLM effectively, especially for more challenged tasks like \textit{ChatIntent} and \textit{ChatStrategy}.

\noindent \textbf{Fine-Tuning Evaluation: } 
Table \ref{tab:finetune_result} exhibits the fine-tuned results of LaHoRe and several 7B baselines. LaHoRe apparently outperforms other baselines, with the DPO version performing even better than SFT. This observation illustrates that generative preference optimization can be another retrieval basis besides the popular contrastive learning, and might be a stronger aligner to hidden rationale-based retrieval.



\noindent \textbf{End-to-End RAG: } 
We finally implement a RAG pipeline with Qwen2-70B-Instruct as the generator and \textit{DailyStrategy} as the test environment. The retrieved reply strategy by LaHoRe is appended into the prompt to enhance LLM generation. Response qualities, including aspects of coherence, fluency and safety, are blindly evaluated by human annotators, and LaHoRe has a 90\% win-rate in the comparison with raw LLM generation.



\begin{table*}[t!]
\centering
\caption{Results of zero-shot ablation experiments. $I_{\text{bc}}$ is the instruction which prompts the retrieval task into a binary-choice question; $D$$\Rightarrow$$Q$ means prepending document before query.
}
\label{tab:ablation_result}
\resizebox{\textwidth}{!}{%
\begin{tabular}{lcc|ccc|ccc|ccc|ccc|c}
\toprule
& \multicolumn{2}{c|}{Dataset($\rightarrow$)} & \multicolumn{3}{c|}{ChatIntent} & \multicolumn{3}{c|}{ChatStrategy} & \multicolumn{3}{c|}{ESconv} & \multicolumn{3}{c|}{PsyQA} & RT$\downarrow$  \\
\cline{2-5} \cline{6-9} \cline{10-12} \cline{13-15}
& $I_{\text{bc}}$ & $D$$\Rightarrow$$Q$ & R@5 & R@10 & p-MRR & R@5 & R@10 & p-MRR & R@1 & R@3 & p-MRR & R@1 & R@3 & p-MRR & (ms) \\ 
\midrule
\parbox[t]{5mm}{\multirow{4}{*}{\rotatebox[origin=c]{90}{
\parbox[c]{1.5cm}{\centering \scriptsize LaHoRe}}}}
& $\times$ & $\times$ & 10.3 & 18.0 & 8.9 & 4.9 & 8.3 & 4.9 & \underline{17.3} & 37.9 & 36.5 & 19.5 & 47.4 & 35.7 & - \\
& $\times$  & \checkmark & 14.3 & 24.7 & 11.0 & 4.7 & 7.6 & 4.8 & 17.2 & 37.0 & 36.5 & 24.9 & 49.5 & 44.5 & - \\
& \checkmark & $\times$  & \bf 31.5 & \bf 50.0 & \underline{19.9} & \underline{10.2} & \underline{21.4} & \underline{8.6} & \bf 18.1 & \bf 43.9 & \bf 39.2 & 30.8 & \underline{52.0} & \underline{48.3} & 47.5 \\
& \checkmark & \checkmark & \underline{30.0} & \underline{48.2} & \bf 22.1 & \bf 16.5 & \bf 28.3 & \bf 11.6 & 16.9 & \underline{42.7} & \underline{38.0} & \bf 32.6 & \bf 52.5 & \textbf{49.5} & \bf 19.3 \\
\bottomrule
\end{tabular}
}
\end{table*}

\begin{figure*}[htbp]
    \centering
    \includegraphics[width=0.25\linewidth]{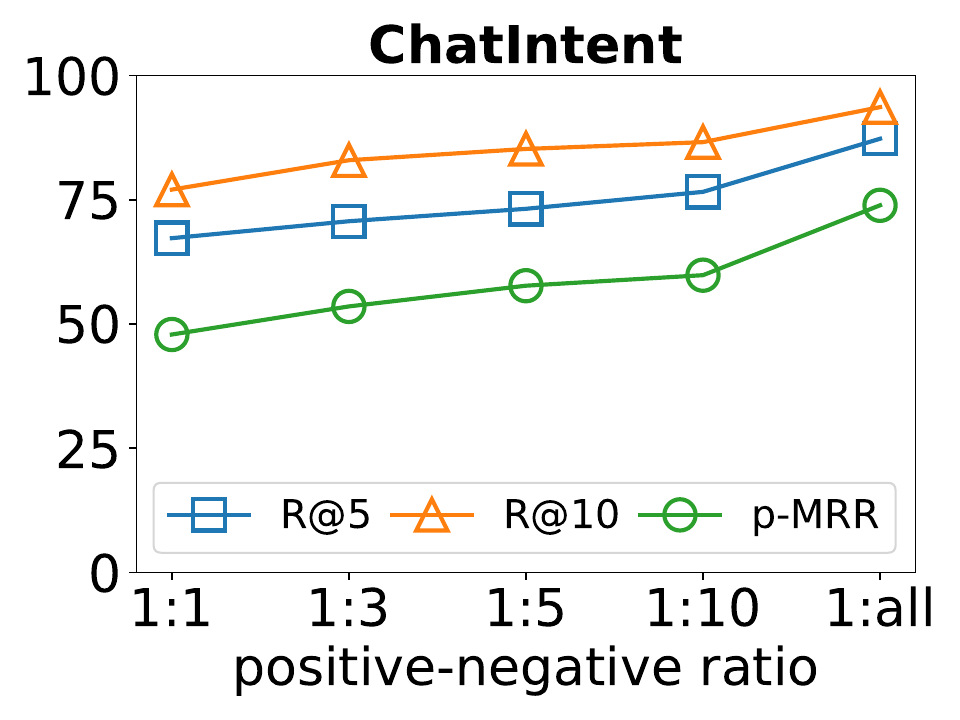}
    \hspace{-0.1in}
    \includegraphics[width=0.25\linewidth]{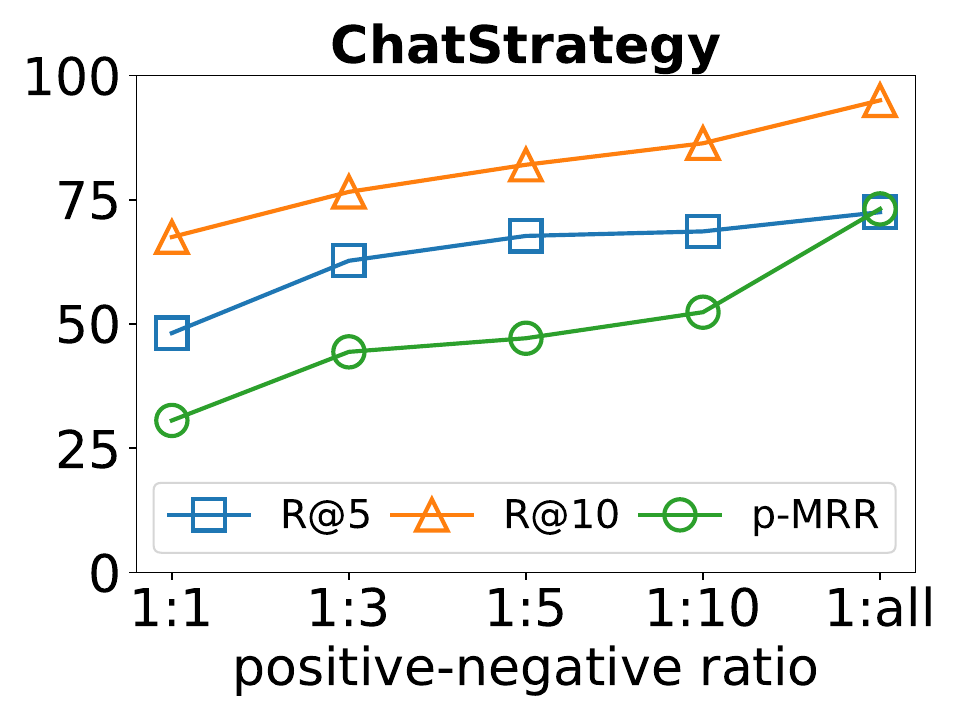}
    \hspace{-0.1in}
    \includegraphics[width=0.25\linewidth]{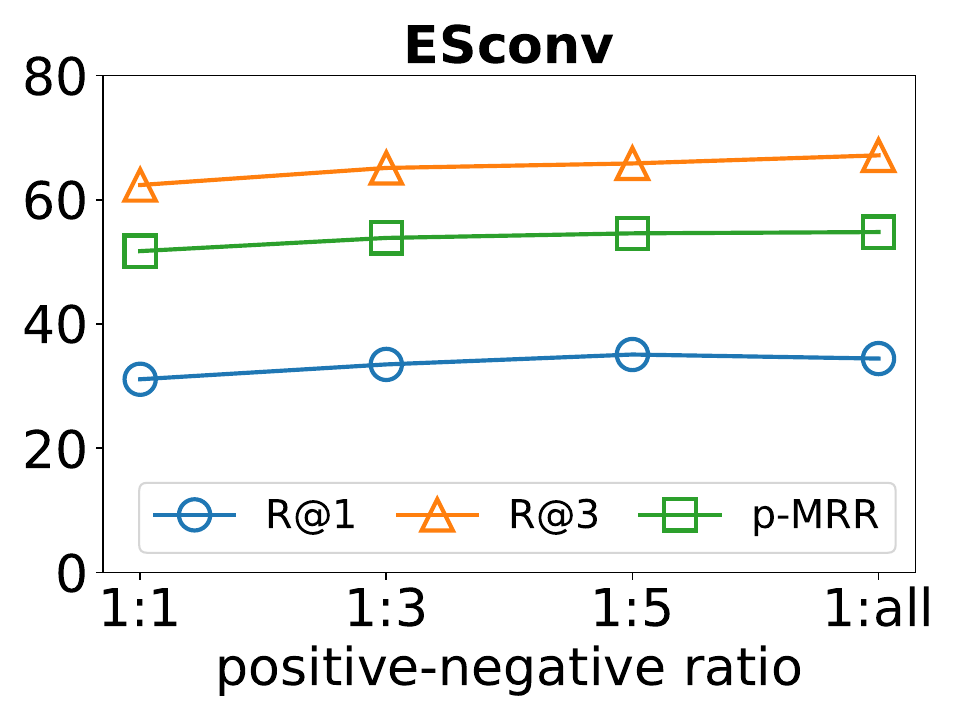}
    \hspace{-0.1in}
    \includegraphics[width=0.25\linewidth]{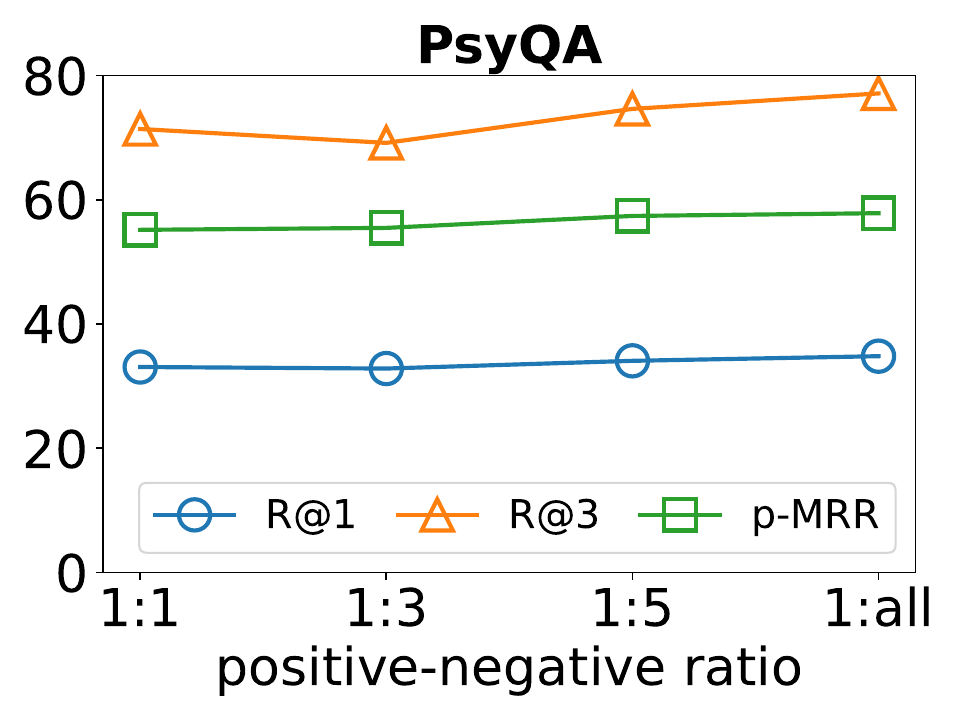}
    \vspace{-0.1in}
    \caption{Retrieval performance plots with different positive-negative ratios.}
    \label{fig:sensitivity}
\end{figure*}

\subsection{Analysis}
\label{sec:analysis}

\noindent \textbf{Ablation Study: } 
We conduct ablation study on two core features, $I_{\text{bc}}$ and $D \Rightarrow Q$. Table \ref{tab:ablation_result}  help answer several important questions:

\noindent \textit{RQ1:} Result of raw LLM (the first row) is worse than not only LaHoRe but also many baselines in Table \ref{tab:zero_shot_result}. This result rules out the capability difference of the basis model.\\
\noindent \textit{RQ2:} With and without $I_{\text{bc}}$ exhibit a huge difference which indicates its effectiveness. \\
\noindent \textit{RQ3:} Performance ablation on $D$$\Rightarrow$$Q$ is relative balanced. However, response time (RT) can be improved apparently with $D$$\Rightarrow$$Q$ where the prefix-cache mechanism applies.


\noindent \textbf{Sensitivity Analysis: } 
In conventional embedding-based retrieval, it is common to construct in-batch negatives or hard negatives and train with contrastive loss. While in our generation-based paradigm, both positive or negative pairs are valid, independent training samples, therefore the positive-negative sample ratio could be an arbitrary choice. Compared with the formal choice of 1:all, we also experiment with other ratios including 1:1, 1:3 and 1.5, with plots of metrics shown in Fig.~\ref{fig:sensitivity}. It is evident that 1:all performs the best, which validates the data-scaling benefit of LaHoRe, even with positive outnumbered by negative samples. Also, this improvement is more apparent when the total number of documents is larger (\textit{ChatIntent} and \textit{ChatStrategy}).

\section{Conclusion}

In this paper, we propose an LLM-based retrieval approach called LaHoRe, to address the challenging hidden rationale retrieval problem. We study the emotional supporting conversation scenario where different reply strategies are retrieved based on the user query. We leverage the reasoning and semantic understanding ability of LLM on this problem with solely generative loss. Specific instruction is designed to prompt LLM to answer a binary-choice question, and prefix decoding technology can be applied to make the retriever computationally efficient. Experimental results verify that both zero-shot and fine-tuned performances of LaHoRe surpass all other baselines on such types of tasks. 


\bibliographystyle{splncs04}
\bibliography{main}

\end{document}